# Size-extensive polarizabilities with intermolecular charge transfer in a fluctuating-charge model

Jiahao Chen (陳家豪) and Todd J. Martínez

*Department of Chemistry, Frederick Seitz Materials Research Laboratory, and the Beckman Institute, University of Illinois at Urbana-Champaign*
*600 South Mathews Avenue, Urbana, IL 61801*

**Abstract**

Fluctuating-charge models have been used to model polarization effects in molecular mechanics methods. However, they overestimate polarizabilities in large systems. Previous attempts to remedy this have been at the expense of forbidding intermolecular charge-transfer. Here, we investigate this lack of size-extensivity and show that the neglect of terms arising from charge conservation is partly responsible; these terms are also vital for maintaining the correct translational symmetries of the dipole moment and polarizability that classical electrostatic theory requires. Also, QTPIE demonstrates linear-scaling polarizabilities when coupling the external electric field in a manner that treats its potential as a perturbation of the atomic electronegativities. Thus for the first time, we have a fluctuating-charge model that predicts size-extensive dipole polarizabilities, yet allows intermolecular charge-transfer.

Polarization is an important but often neglected phenomenon in many classical molecular dynamics simulations.[1-3] Of the methods invented to model polarization in this context,[4, 5] fluctuating-charge models provide a unified treatment of polarization and charge transfer, and thus most strongly resemble quantum-mechanical electronic structure methods. While various implementations like EEM [6, 7], QEq [8, 9], *fluc*-q [10-14], AACT [15-17], CPE [18-20], ABEEM [21, 22], CHARMM C22 [23-25], EVB-based models[26-28], and others[29-48] are popular, they still suffer from some uncorrected pathologies. For example, many fluctuating-charge models are known to:

1. describe physically unreasonable charge distributions for geometries far from equilibrium,[40, 49-52]

2. exhibit super-linear scaling of polarizabilities unless intermolecular charge transfer is forbidden,[15-17, 19, 53, 54] and

3. predict the wrong direction of charge transfer.[40]

These problems are not due to imperfect parameters, but are symptomatic of deficiencies in the theoretical framework. Our previous work[50, 51] had led us to propose the QTPIE charge model,[49, 55] which has already been shown to cure the first problem for systems are electrically neutral overall. Here, we describe a solution to the last two problems as well.

In fluctuating-charge models, the electrostatic energy usually takes the form

$$E(\mathbf{q}) = \sum_{i=1}^{N} q_i v_i + \frac{1}{2} \sum_{i,j=1}^{N} q_i q_j J_{ij} \equiv \mathbf{q} \cdot \mathbf{v} + \frac{1}{2} \mathbf{q}^T \mathbf{J} \mathbf{q} \qquad (1)$$

where $q_i$ is the charge on atom $i$, $J_{ij}$ is the Coulomb interaction between atoms $i$ and $j$, $J_{ii}$ is the chemical hardness of atom $i$, and $v_i$ is the electronegativity of atom $i$. Due to electronegativity equalization,[56, 57] the solution is constrained to a total charge $Q = \mathbf{q} \cdot \mathbf{1} \equiv \sum_{i=1}^{N} q_i$. By introducing the chemical potential[58, 59] $\mu$ as a Lagrange multiplier, we construct the free energy $F(\mathbf{q}, \mu) = E - \mu(\mathbf{q} \cdot \mathbf{1} - Q)$ and minimize it to obtain

$$\begin{pmatrix} \mathbf{J} & \mathbf{1} \\ \mathbf{1}^T & 0 \end{pmatrix} \begin{pmatrix} \mathbf{q} \\ \mu \end{pmatrix} = \begin{pmatrix} -\mathbf{v} \\ Q \end{pmatrix} \tag{2}$$

Gaussian elimination on $\mu$ gives the analytic solution

$$\begin{pmatrix} \mathbf{q} \\ \mu \end{pmatrix} = \sigma \begin{pmatrix} \sigma^{-1} \mathbf{J}^{-1} - (\mathbf{J}^{-1} \mathbf{1})(\mathbf{J}^{-1} \mathbf{1})^T & \mathbf{J}^{-1} \mathbf{1} \\ (\mathbf{J}^{-1} \mathbf{1})^T & -1 \end{pmatrix} \begin{pmatrix} -\mathbf{v} \\ Q \end{pmatrix} = \begin{pmatrix} -\mathbf{J}^{-1} \mathbf{v} + \mu \mathbf{J}^{-1} \mathbf{1} \\ -\sigma(\mathbf{1}^T \mathbf{J}^{-1} \mathbf{v} + Q) \end{pmatrix} \tag{3}$$

where $\sigma^{-1} = \mathbf{1}^T \mathbf{J}^{-1} \mathbf{1}$ is the Schur complement of $\mathbf{J}$ in Eq. (2). Therefore, the electrostatic energy attains its minimum at

$$E_0 = -\tfrac{1}{2} \mathbf{v}^T \mathbf{J}^{-1} \mathbf{v} + \tfrac{1}{2} \sigma (\mathbf{1}^T \mathbf{J}^{-1} \mathbf{v} + Q)^2 \tag{4}$$

Interestingly, the scalar $\sigma$, which has dimensions of hardness, quantifies how much the charge constraint changes the charge distribution.

Consider models where $\mathbf{v}$ is a vector of atomic electronegativities, i.e. $v_i = \chi_i$. We then couple such a model to an external electrostatic field $\mathrm{E}^\nu$ (where $\nu$ indexes spatial directions) by introducing the usual dipole coupling term so that the electrostatic energy becomes

$$E(\mathbf{q}, \mathrm{E}^\nu) = E(\mathbf{q}) - \mathbf{q} \cdot \sum_\nu \mathbf{R}_\nu \mathrm{E}^\nu = \mathbf{q} \cdot \left( \mathbf{v} - \sum_\nu \mathbf{R}_\nu \mathrm{E}^\nu \right) + \tfrac{1}{2} \mathbf{q}^T \mathbf{J} \mathbf{q} \tag{5}$$

Thus the effect of coupling to an electrostatic field is to transform the atomic electronegativities by

$$\chi_i \mapsto \chi_i - \sum_\nu R_{i\nu} \mathrm{E}^\nu \tag{6}$$

The dipole moment and polarizability tensor are then

$$d_\nu = \frac{\partial E_0}{\partial \mathrm{E}^\nu} = \mathbf{R}_\nu^T \mathbf{J}^{-1}\left(\mathbf{v} - \sum_\lambda \mathbf{R}_\lambda^T \mathrm{E}^\lambda\right) - \sigma\left(\mathbf{1}^T \mathbf{J}^{-1} \mathbf{R}_\nu\right)\left(\mathbf{1}^T \mathbf{J}^{-1} \mathbf{v} + Q\right) \tag{7}$$

$$\alpha_{\nu\lambda} = \frac{\partial d_\nu}{\partial \mathrm{E}^\lambda} = -\mathbf{R}_\nu^T \mathbf{J}^{-1} \mathbf{R}_\lambda + \sigma\left(\mathbf{1}^T \mathbf{J}^{-1} \mathbf{R}_\nu\right)\left(\mathbf{1}^T \mathbf{J}^{-1} \mathbf{R}_\lambda\right) \tag{8}$$

For all systems of finite extent, $\sigma > 0$; however, many published results in the literature are, surprisingly, missing the terms in $\sigma$. This leads to physically incorrect behavior. For example, under the change of origin $R_{i\nu} \mapsto R_{i\nu} + \delta_\nu$, the dipole moment transforms as $d_\nu \mapsto d_\nu + \delta_\nu Q$ and the polarizability as $\alpha_{\nu\lambda} \mapsto \alpha_{\nu\lambda}$ if and only if terms in $\sigma$ are included. It is therefore unnecessary to require specific choices of origin[23-25, 32, 54] to simulate translational invariance.

We now investigate the size extensivity of (7) and (8). Consider a system with $n$ identical copies of a subsystem comprised of $m$ atoms, with each copy separated by a distance $\Delta_\nu$ that is larger than the spatial extent of one subsystem. We use the overbar to denote quantities related to a single subsystem. The nuclear coordinates are then

$$\mathbf{R}_\nu = \begin{pmatrix} \overline{\mathbf{R}}_\nu \\ \overline{\mathbf{R}}_\nu + \Delta_\nu \overline{\mathbf{1}} \\ \vdots \\ \overline{\mathbf{R}}_\nu + \Delta_\nu(n-1)\overline{\mathbf{1}} \end{pmatrix} = \begin{pmatrix} \overline{\mathbf{R}}_\nu \\ \overline{\mathbf{R}}_\nu \\ \vdots \\ \overline{\mathbf{R}}_\nu \end{pmatrix} + \Delta_\nu \begin{pmatrix} \overline{\mathbf{0}} \\ \overline{\mathbf{1}} \\ \vdots \\ (n-1)\overline{\mathbf{1}} \end{pmatrix} \tag{9}$$

and $\mathbf{v} = \left(\overline{\mathbf{v}}, ..., \overline{\mathbf{v}}\right)^T$. In the limit $|\Delta| \to \infty$, the subsystems decouple and $\mathbf{J}$ becomes approximately block diagonal, with inverse

$$\mathbf{J}^{-1} = \begin{pmatrix} \bar{\mathbf{J}}^{-1} & \mathbf{0} & \cdots & \mathbf{0} \\ \mathbf{0} & \bar{\mathbf{J}}^{-1} & \ddots & \vdots \\ \vdots & \ddots & \ddots & \mathbf{0} \\ \mathbf{0} & \cdots & \mathbf{0} & \bar{\mathbf{J}}^{-1} \end{pmatrix} + O\left(\left|\Delta\right|^{-1}\right) \tag{10}$$

In this limit, the total dipole moment and polarizability then become

$$d_\nu = n\bar{d}_\nu + \tfrac{1}{2}(n-1)(n-2)\bar{Q}\Delta_\nu + O\left(\Delta_\nu / |\Delta|\right) \tag{11}$$

$$\alpha_{\nu\lambda} = n\bar{\alpha}_{\nu\lambda} - \frac{(n-1)(n-2)(n^2-3n-6)}{12n}\Delta_\nu\Delta_\lambda\bar{\sigma} + O\left(\frac{\Delta_\nu\Delta_\lambda}{|\Delta|}\right) \tag{12}$$

where the subsystem dipole moment and polarizability are defined analogously to (7) and (8), i.e.

$$\bar{d}_\nu = \bar{\mathbf{R}}_\nu^T \bar{\mathbf{J}}^{-1}\left(\bar{\mathbf{v}} - \sum_\lambda \bar{\mathbf{R}}_\lambda^T \mathbf{E}^\lambda\right) - \bar{\sigma}\left(\bar{\mathbf{1}}^T \bar{\mathbf{J}}^{-1} \bar{\mathbf{R}}_\nu\right)\left(\bar{\mathbf{1}}^T \bar{\mathbf{J}}^{-1}\bar{\mathbf{v}} + \bar{Q}\right) \tag{13}$$

$$\bar{\alpha}_{\nu\lambda} = -\bar{\mathbf{R}}_\nu^T \bar{\mathbf{J}}^{-1}\bar{\mathbf{R}}_\lambda + \bar{\sigma}\left(\bar{\mathbf{1}}^T \bar{\mathbf{J}}^{-1}\bar{\mathbf{R}}_\nu\right)\left(\bar{\mathbf{1}}^T \bar{\mathbf{J}}^{-1}\bar{\mathbf{R}}_\lambda\right) \tag{14}$$

where $\bar{\sigma}^{-1} = \bar{\mathbf{1}}^T \bar{\mathbf{J}}^{-1}\bar{\mathbf{1}}$ and $\bar{Q} = Q / n$ is the total charge of each identical subsystem. The second term in (11) represents the summed contributions of $m$ point charges, each of charge $\bar{Q}$ and placed at coordinates $\bar{\mathbf{0}}, \bar{\mathbf{1}}, ...,(n-1)\bar{\mathbf{1}}$ respectively. When $Q = 0$, the dipole moment (11) becomes size-extensive. However, the second term in the polarizability expression (12) grows cubically with $n$, which is physically incorrect. Note that neither the dipole moment nor the polarizability would be size consistent if the terms in $\bar{\sigma}$ were neglected.

We now describe a way to obtain size-extensive dipole polarizabilities in QTPIE. As shown earlier[52, 55], any bond-space fluctuating charge model[32, 60], including QTPIE, can be written in the form of (1) but with effective atomic voltages $\mathbf{v}$ that are not identical to the atomic electronegativities. For QTPIE, the model is defined only for $Q =$

0, but this is not a serious limitation in practice. The effective atomic voltages are given by

$$v_i = \tau_i \sum_{j=1}^{N} S_{ij} \left( \chi_i - \chi_j \right) \quad (15)$$

where $S_{ij} = \left\langle \phi_i \middle| \phi_j \right\rangle$ is the overlap integral between atomic basis functions $\left\{ \phi_i \left( r; R_i \right) \right\}$ and $\tau_i^{-1} = \sum_{k=1}^{N} S_{ik}$. Instead of the usual coupling (5), we now apply the transformation (6) directly to (15), so that

$$v_i = \tau_i \sum_{j=1}^{N} S_{ij} \left( \chi_i - \chi_j - \sum_{v} \left( R_{iv} - R_{jv} \right) E^v \right) \quad (16)$$

In models where $v_i = \chi_i$, this prescription is identical to (5). However, in QTPIE, only (16) treats the external electrostatic potential $-\sum_{v} R_{iv} E^v$ on the $i^{\text{th}}$ atom on the same footing as the internal potential $\chi_i$. The dipole moment and polarizability tensor are then

$$d_v = \sum_{i,i',j,j'=1}^{N} \tau_i \tau_{i'} S_{ij} S_{i'j'} \left( R_{iv} - R_{jv} \right) \left( \sigma \left( \mathbf{1}^T \mathbf{J}^{-1} \right)_i \left( \mathbf{1}^T \mathbf{J}^{-1} \right)_{i'} - \left( \mathbf{J}^{-1} \right)_{ii'} \right)$$
$$\times \left( \chi_{i'} - \chi_{j'} - \sum_{\lambda} \left( R_{i'\lambda} - R_{j'\lambda} \right) E^\lambda \right) \quad (17)$$

$$\alpha_{v\lambda} = \sum_{i,i',j,j'=1}^{N} \tau_i \tau_{i'} S_{ij} S_{i'j'} \left( R_{iv} - R_{jv} \right) \left( \sigma \left( \mathbf{1}^T \mathbf{J}^{-1} \right)_i \left( \mathbf{1}^T \mathbf{J}^{-1} \right)_{i'} - \left( \mathbf{J}^{-1} \right)_{ii'} \right) \left( R_{i'\lambda} - R_{j'\lambda} \right) \quad (18)$$

Note that (17) and (18) still retain the correct translational invariance.

When we apply the subsystem decomposition of (9) and (10) to (17) and (18), the overlap matrix element decays exponentially quickly with interatomic distance and thus attenuates inter-subsystem interactions; the effective atomic voltages (15) become

$$\bar{v}_i = \sum_{j=1}^{m} S_{ij}\left(\chi_i - \chi_j\right) / \left(\sum_{k=1}^{m} S_{ik}\right) + O\left(e^{-|\Delta|}\right).$$ Then

$$d_v = n\bar{d}_v + O\left(e^{-|\Delta|}\right) \tag{19}$$

$$\alpha_{v\lambda} = n\bar{\alpha}_{v\lambda} + O\left(n^3 e^{-|\Delta|}\right) \tag{20}$$

Therefore, the overlaps give rise to size-extensivity. Importantly, this does not come at the price of forbidding intermolecular charge transfer *a priori*; unlike previously proposed topological solutions to the size-extensivity problem.[10-14, 54, 61, 62]

For illustrative purposes, we now use (17) and (18) to calculate electrostatic properties of linear coplanar water chains consisting of one through 25 water molecules. To better study the size extensivity, we use idealized geometries instead of optimized geometries for each chain. The oxygen atoms are collinear and spaced 2.870 Å apart; the hydrogen atoms are all coplanar with transverse separations of 1.514 Å and with O–H bond lengths of 1.000 Å. We compare our results to those from QEq and from *ab initio* calculations employing density-fitted local MP2[63] with Dunning's aug-cc-pVTZ basis set[64].

As expected, QTPIE predicts correctly size-extensive  dipole moments (Figure 1), transverse in-plane polarizabilities (Figure 2), and longitudinal polarizabilities (Figure 3). In contrast, QEq predicts superlinear scaling of the longitudinal polarizability as shown in Figure 3. Despite our use of non-optimized parameters in QTPIE - having set them to those of QEq - there is remarkably good agreement between QTPIE and the *ab initio* data.

Furthermore, this size-extensivity is observed in the presence of intermolecular charge transfer. Figure 4 shows the total molecular charge on each water molecule in the 15-water chain as calculated using QEq, QTPIE and Mulliken population analysis of the *ab initio* wavefunction calculated above. Surprisingly, QEq predicts intermolecular charge transfer in the opposite direction as expected from chemical intuition and the *ab initio* data. In contrast, QTPIE predicts the correct direction of charge flow

In conclusion, we have shown that the correct solution of fluctuating-charge models contains terms that have been previously neglected. This neglect leads to unphysical violations of the translational properties of dipole moments and polarizabilities, and are partly responsible for the apparent lack of size extensivity. While including these terms lead to size-extensive dipoles, they still do not cure the superlinear scaling of polarizabilities. To solve this, we propose the coupling (16). With it, QTPIE predicts linear-scaling polarizabilities while allowing intermolecular charge transfer. Intermolecular charge transfer between coplanar, linear water chains in QTPIE is retained without any adjustment of the original QEq parameters. Thus we have shown that QTPIE is a fluctuating charge model that is useful for accounting for both polarization and charge transfer effects in general systems.

**Acknowledgment.** This work was supported by DOE DE-FG02-05ER46260.

# List of figures

Figure 1. Dipole moment of planar water chains as a function of the number of water molecules in the chain in Debyes, as calculated by QEq (red dashed line), QTPIE (black solid line) and an *ab initio* method (density-fitted local MP2 with Dunning aug-cc-pVTZ basis sets, blue dot-dashed line). Water chain geometries are described in the main text.

Figure 2. Transverse in-plane component of the dipole polarizability of planar water chains as a function of the number of water molecules in the chain in cubic Ångströms, as calculated by QEq (red dashed line), QTPIE (black solid line) and an *ab initio* method (density-fitted local MP2 with Dunning aug-cc-pVTZ basis sets, blue dot-dashed line). This transverse component is parallel to the H–H axes. Water chain geometries are described in the main text.

Figure 3. Longitudinal component of the dipole polarizability of planar water chains as a function of the number of water molecules in the chain in cubic Ångströms, as calculated by QEq (red dashed line), QTPIE (black solid line) and an *ab initio* method (density-fitted local MP2 with Dunning aug-cc-pVTZ basis sets, blue dot-dashed line). The longitudinal component is parallel to the O–O axis. Water chain geometries are described in the main text.

Figure 4. Total molecular charge on water molecules in a planar linear chain of 15 water molecules, as calculated QEq (red dashed line), QTPIE (black solid line) and Mulliken population analysis of the wavefunction from density-fitted local MP2 with Dunning aug-

cc-pVTZ basis sets (blue dot-dashed line). Note the incorrect direction of charge transfer in QEq. Water chain geometries are described in the main text.

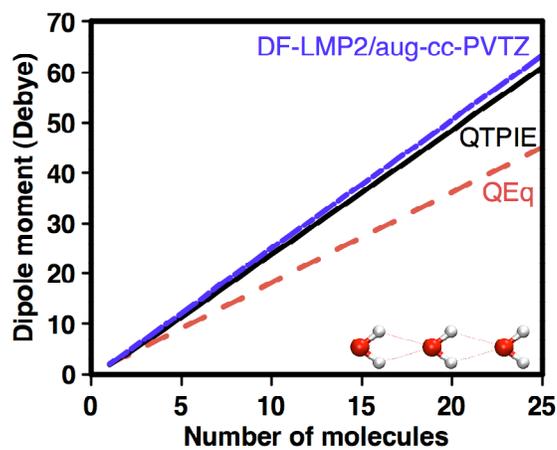

**Figure 1.** Dipole moment of planar water chains as a function of the number of water molecules in the chain in Debyes, as calculated by QEq (red dashed line), QTPIE (black solid line) and an *ab initio* method (density-fitted local MP2 with Dunning aug-cc-pVTZ basis sets, blue dot-dashed line). Water chain geometries are described in the main text.

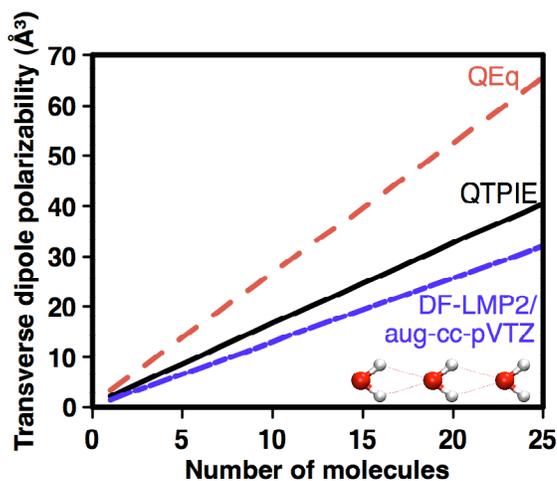

**Figure 2.** Transverse in-plane component of the dipole polarizability of planar water chains as a function of the number of water molecules in the chain in cubic Ångströms, as calculated by QEq (red dashed line), QTPIE (black solid line) and an *ab initio* method (density-fitted local MP2 with Dunning aug-cc-pVTZ basis sets, blue dot-dashed line). This transverse component is parallel to the H–H axes. Water chain geometries are described in the main text.

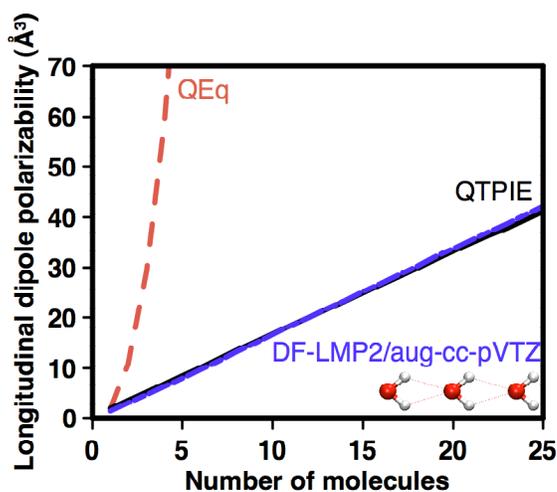

**Figure 3.** Longitudinal component of the dipole polarizability of planar water chains as a function of the number of water molecules in the chain in cubic Ångströms, as calculated by QEq (red dashed line), QTPIE (black solid line) and an *ab initio* method (density-fitted local MP2 with Dunning aug-cc-pVTZ basis sets, blue dot-dashed line). The longitudinal component is parallel to the O–O axis. Water chain geometries are described in the main text.

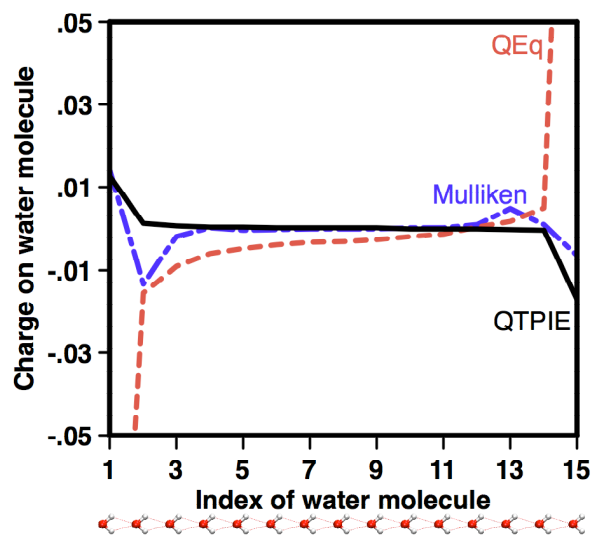

**Figure 4.** Total molecular charge on water molecules in a planar linear chain of 15 water molecules, as calculated QEq (red dashed line), QTPIE (black solid line) and Mulliken population analysis of the wavefunction from density-fitted local MP2 with Dunning aug-cc-pVTZ basis sets (blue dot-dashed line). Note the incorrect direction of charge transfer in QEq. Water chain geometries are described in the main text.